\documentclass[usenatbib,12pt,preprint]{aastex}
\slugcomment{Accepted 31 August 2007}
\usepackage{graphicx}
\shorttitle{Pflamm-Altenburg, Weidner and Kroupa}
\shortauthors{Converting H$\alpha$ luminosities into SFRs}

\begin{document}
\title{Converting H$\alpha$ luminosities into star formation rates}
\author{Jan~Pflamm-Altenburg}
\affil{Argelander-Institut f\"ur Astronomie (AIfA)}
\affil{Auf dem H\"ugel 71, 53121 Bonn, Germany}
\email{jpflamm@astro.uni-bonn.de}
\author{Carsten Weidner}
\affil{Departamento de Astronom{\'i}a y Astrof{\'i}sica, Pontificia
  Universidad Cat{\'o}lica de Chile}
\affil{Av. Vicu{\~n}a MacKenna 4860,
      Macul, Santiago, Chile}
\email{cweidner@astro.puc.cl}
\and
\author{Pavel Kroupa}
\affil{Argelander-Institut f\"ur Astronomie (AIfA)}
\affil{Auf dem H\"ugel 71, 53121 Bonn, Germany}
\email{pavel@astro.uni-bonn.de}

\begin{abstract}
Star-formation rates (SFRs) of galaxies are commonly calculated by converting
the measured H$\alpha$ luminosities ($L_\mathrm{H\alpha}$) into current SFRs.
This conversion is based on a constant initial mass function (IMF) independent 
of the total SFR. 
As recently recognised the maximum stellar mass in a star cluster is limited
by the embedded total cluster mass and, in addition,  
the maximum embedded star cluster
mass is constrained by the current SFR. The combination
of these two relations leads to an integrated
galaxial initial stellar mass function (IGIMF, the IMF for the whole
galaxy) which is steeper in the 
high mass regime than the constant canonical IMF, and is dependent on
the SFR of the galaxy.
Consequently, the $L_\mathrm{H\alpha}$-SFR relation becomes non-linear
and flattens  for low SFRs. Especially for 
dwarf galaxies the SFRs can 
be underestimated by up to three orders of magnitude.
We revise the existing linear $L_{\mathrm{H}\alpha}$-SFR relations 
using our IGIMF notion. These are likely
to lead to a revision of the cosmological star formation histories.
We also demonstrate that in the case of the Sculptor 
dwarf irregular galaxies the IGIMF-formalism implies  
a linear dependence of the total SFR on the total galaxy gas mass.
A constant gas depletion time scale of a few Gyrs results
independently of the galaxy gas mass with a reduced scatter compared to the 
conventional results. Our findings are qualitatively independent of the
explicit choice of the IGIMF details and challenges current star formation
theory in dwarf galaxies.  

\end{abstract}
\section{Introduction}
Low-mass dwarf galaxies are very important objects to test the theoretical
understanding of star formation, galactic evolution and cosmology.
One key issue in this process is the accurate determination 
of current SFRs. Because massive stars have short life times
the number of currently existing massive stars determines the actual 
massive star formation rate. The number of massive stars is commonly
obtained from the integrated  H$\alpha$ luminosity of the target galaxy,
after correcting the observations for extinction and N {\sc ii} emission.
If the fraction of massive star formation of  the total star 
formation is known then the meassured integrated H$\alpha$ luminosity
of a galaxy can be converted into its current SFR.

The standard method is to apply one initial mass function (IMF)
to  the entire
population of newly formed stars in the whole galaxy. 
This standard method provides a linear relation between the integrated
H$\alpha$ luminosity and the current SFR 
\citep{kennicutt1983a,kennicutt1994a,kennicutt1998a}. Originally used
for normal disk galaxies such linear  relations have also been 
applied to dwarf irregular galaxies \citep{skillman2003a}.

But even if star formation in each young 
star cluster follows the same canonical IMF
\citep{kroupa2001a,kroupa2002a,pflamm_altenburg2006a},
the total stellar population of newly formed stars of all young star
clusters together follows a distribution function which is steeper than the
canonical IMF in the high mass regime \citep{weidner2003a,weidner2005a}. This
distribution function is called the integrated galaxial initial stellar mass
function (IGIMF) and deviates increasingly from the underlying canonical IMF
with decreasing galactic-wide star formation and total galaxy mass
\citep{weidner2005a}. 

This effect has undoubtedly consequences for the cosmological evolution
of the number of supernovae per low-mass star and the chemical enrichment
of galaxies of different mass. As a first step  
the formulation of the IGIMF has been 
successfully used to naturally explain the mass-metallicity relation of galaxies
\citep*{koeppen2007a}. 

Observational evidence supporting this IGIMF concept has been published
by \citet{selman2005a} who find a steeper IMF-slope of the high-mass regime
in the field population of 30~Doradus than in the cluster NGC~2070
as predicted by \citet{weidner2003a}. 
Note that the \citet{scalo1986a} Milky-Way-disc IMF index 
$\alpha\approx 2.7$ \citep*{kroupa1993a} is  larger than the canonical IMF 
index $\alpha \approx 2.35$. This difference is a natural consequence
of the IGIMF notion \citep{weidner2003a}. Using the Scalo index  
\citet{diehl2006a} find a consistency of the supernova rate derived
from the Galactic $^{26}$Al gamma-ray flux and deduced from a survey
from local O and B stars, extrapolated to the whole Galaxy
\citep{mckee1997a}.
Noteworthy is the finding
by  \citet{lee2004a} that LSB galaxies have much steeper than Salpeter
massive-star IMFs than Milky-Way-type galaxies, as expected from the
IGIMF formalism. 
\citet{hoversten2007a}
find evidence for a non-universal stellar IMF from the integrated 
properties of SDSS galaxies. Bright galaxies have a high-massive star
slope of $\approx 2.4$, whereas fainter galaxies prefer steeper IMFs.
The observed effect is very much in-line with the theoretical predictions
of \citet{weidner2005a}.

For a given global SFR the total number of massive stars based 
on an IGIMF is less than in the standard procedure. Thus, the expected
IGIMF-H$\alpha$ luminosity is less than the H$\alpha$ luminosity calculated
in the standard way, implying that SFRs  of galaxies
determined by the standard method are 
systematically underestimated.     

After  a summary of the IGIMF basics  
we develop for different IGIMF-models 
a relation between the SFR and the produced H$\alpha$
luminosity and provide easy-to-use fitting functions 
(Sec. \ref{sec_SFR_Halpha}).
In Sec. \ref{sec_sfr_dIrr} we apply our $L_{\mathrm{H}\alpha}$-SFR
relation to the dwarf irregular galaxies in the Sculptor cluster
and discuss the changes of some of their parameters.
Finally, we explore the detection limits for star formation 
(Sec. \ref{sec_invisible_sf}). 

\section{SFR-L$_{\mathrm{H}\alpha}$-relation}
\label{sec_SFR_Halpha}
For a given SFR  we assume that star formation takes place 
in star clusters distributed according to  a mass function
which is called the embedded cluster mass function (ECMF)
of stellar masses (i.e. the function of the stellar-mass content in 
clusters before they expel their 
residual gas), 
$\xi_\mathrm{ecl}(M_\mathrm{cl})$. 
These star clusters are formed within a time-span, $\delta t$,
which is called the ''star-formation epoch'' of the galaxy
\citep{weidner2005a}.

The total mass of all formed stars, $M_\mathrm{SFR}$, 
in all star clusters within the epoch 
$\delta t$ is then given by 
\begin{equation}
  \label{M_SFR}
  M_\mathrm{SFR}=\mathrm{SFR} \;\;\delta t = \int_{M_\mathrm{ecl,min}}^{M_\mathrm{ecl,max}}
  M_\mathrm{ecl}\;\xi_\mathrm{ecl}(M_\mathrm{ecl})\;\mathrm{d}M_\mathrm{ecl}\;\;\;,
\end{equation}
where $M_\mathrm{ecl, min}$ and $M_\mathrm{ecl, max}$ are, respectively,
the least-massive and most-massive embedded cluster able to form.
The total ensemble of all  newly formed  stars has a mass spectrum 
described by the IGIMF, $\xi_\mathrm{IGIMF}(m)$, constructed below. 
A star with the mass $m$
produces a total number, $N_{\mathrm{ion},\delta t}(m)$, of ionising photons
during the epoch $\delta t$. The total number of ionising photons emitted
by all young stars within $\delta t$ is given by 
\begin{equation}
  N_{\mathrm{ion}, \delta t} =
  \int_{m_\mathrm{low}}^{m_\mathrm{max}}\xi_\mathrm{IGIMF}(m)
  \;N_{\mathrm{ion},\delta t}(m)\;\mathrm{d}m\;\;\;,
\end{equation}   
where $m_\mathrm{low}$ and $m_\mathrm{max}$ are, respectively, the minimum
and maximum stellar mass. 
The calculation of $N_{\mathrm{ion}, \delta t}$ will be described in
  Sec.~\ref{ionising_photons}.

Assuming that a fraction, $\mu$, of all ionising photons
leads to H$\alpha$ emission in the surrounding gas then the
resulting H$\alpha$ luminosity is 
\begin{equation}
  L_\mathrm{H\alpha} = \mu \;3.0207\cdot 10^{-12}\;\mathrm{erg} 
  \;N_\mathrm{ion} / \delta t\;\;\;,
\end{equation}
where the energy of one H$\alpha$-photon is $3.0207\cdot 10^{-12}$~erg.
Throughout this paper we assume $\mu = 1$. 
It can be argued that the time-span $\delta t$ can be removed in this
formulation by considering rates directly. But the total mass, $M_\mathrm{SFR}$,
defined by Eq. \ref{M_SFR} and depending explicitly on $\delta t$,
is an upper limit of the ECMF, as no cluster
can be formed more massive than the available material. 

\subsection{IGIMF}
\label{IGIMF}
\citet{larsen2000a,larsen2002a,larsen2002b} and \citet{larsen2000b} 
found that the V-band luminosity of the brightest young cluster correlates with
the global SFR. 
Based on the conclusion by \citet{larsen2002a} 
that the so-called super-clusters are 
just the young and massive upper end of a cluster mass function, 
\citet{weidner2004b} derived a relation between the maximum embedded
cluster mass, $M_\mathrm{ecl,max}$, in a galaxy and the current global SFR
\begin{equation}
  \label{eqn_m_ecl_max_sfr}
  \log_{10}\left(M_\mathrm{ecl,max}\right) =
  \log_{10}\left(k_\mathrm{ML}\right)
  + 0.75 \cdot \log_{10}\left(\mathrm{SFR}\right)+6.77\;\;\;,
\end{equation}
where $k_\mathrm{ML}$ is the mass-to-light ratio, typically 0.0144 for
young ($<$~6~Myr) stellar populations. They showed that this relation
can be reproduced theoretically if the ECMF is a power-law with
a slope of 2.35 and the 
entire star cluster population ranging from 
$M_\mathrm{ecl,min}\approx$~5~M$_\odot$ to $M_\mathrm{ecl,max}$ is born
within
a time-span of $\delta t\approx$~10~Myr, independently of the SFR.
The concept of a relation between the SFR of a galaxy and 
the mass of the most massive star cluster has been developed further 
to derive the star formation history of galaxies \citep{maschberger2007a}. 
 
Each cluster with the mass $M_\mathrm{ecl}$ then forms $N$ stars
between the mass limits $m_\mathrm{low}$ and $m_\mathrm{max}$ according
to the canonical IMF,
\begin{equation}
  N = \int_{m_\mathrm{low}}^{m_\mathrm{max}}\xi_{M_\mathrm{ecl}}(m) 
  \mathrm{d}m\;\;\;.
\end{equation} 
This canonical IMF, $\xi_{M_\mathrm{ecl}}(m) $, is a two-part-power law
and is here calculated using the algorithm 
by \citet{pflamm_altenburg2006a}. The normalisation constant of the
IMF in an 
individual cluster, $\xi_{M_\mathrm{ecl}}(m)$, with the mass $M_\mathrm{ecl}$ and the maximum stellar mass, $m_\mathrm{max}$, 
in this star cluster are determined by the two equations
\citep{weidner2004a}
\begin{equation}
  \label{equ_m-m1}
  M_\mathrm{ecl} = \int_{m_\mathrm{low}}^{m_\mathrm{max}} m \;\xi_{M_\mathrm{ecl}}(m) 
  \mathrm{d}m\;\;\;,
\end{equation}
\begin{equation}
  \label{equ_m-m2}
  1 = \int_{m_\mathrm{max}}^{m_\mathrm{max*}} \xi_{M_\mathrm{ecl}}(m) 
  \mathrm{d}m\;\;\;.
\end{equation}
The existence of an upper physical stellar mass, $m_\mathrm{max*}$,
of about 150~M$_\odot$ has been 
found in statistical examinations of several clusters 
\citep{weidner2004a,figer2005a,oey2005a,koen2006a,maiz_apellaniz2006a}. 
The canonical IMF,
\begin{equation}
  \xi_{M_\mathrm{ecl}}(m) = k m^{-\alpha_\mathrm{i}}\;\;\;,
\end{equation}
has the slopes
\begin{equation}
  \label{equ_imf_slopes}
  \begin{array}{l@{\;\;\;,\;\;\;}l}
    \alpha_1 = +1.30&0.08 \le m/M_\odot \le 0.50\\
    \alpha_2 = +2.35&0.50 \le m/M_\odot \le 1.00\\
    \alpha_3 = +2.35&1.00 \le m/M_\odot \le m_\mathrm{max}\\
  \end{array}\;\;\;,
\end{equation}
as used by \citet{weidner2005a} who constructed different
IGIMF-models. The third slope above 1~M$_\odot$ has been introduced 
for varying the IMF-slope and constructing non-canonical models
(as in Tab.~\ref{tab_igimf_scenarios}).

\begin{figure}
  \plotone{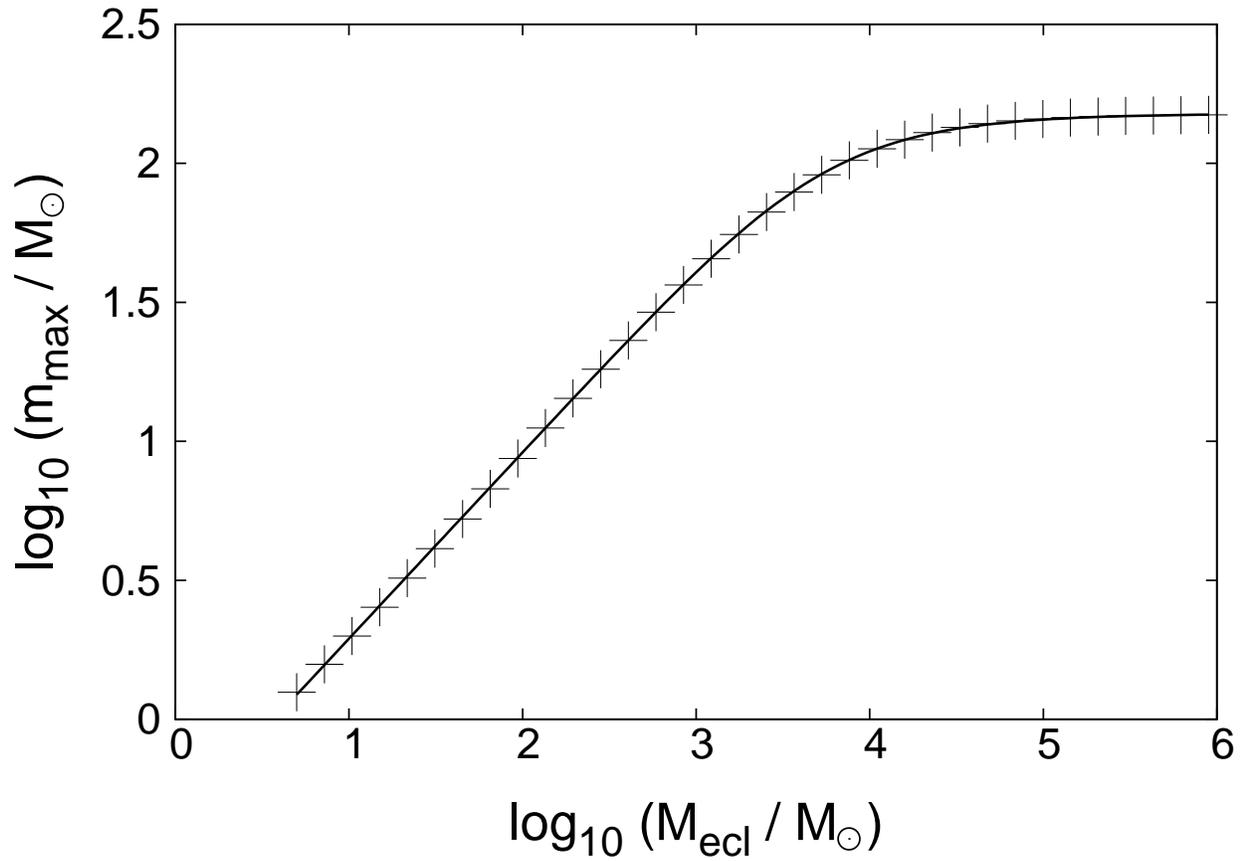}
  \caption{The $m_\mathrm{max}$-$M_\mathrm{ecl}$ relation defined by
    Eqs. \ref{equ_m-m1} and \ref{equ_m-m2} with 
    $m_\mathrm{max*}=150\;\mathrm{M}_\odot$ and fitted by Eq. \ref{equ_m-m}
    (solid curve).
    Only each 30th point of the numerical solution is marked by a cross.
  }
  \label{fig_m-m}
\end{figure}

The pair of the two integral expressions 
(Eq.~\ref{equ_m-m1},~\ref{equ_m-m2}) can not be solved 
analytically to obtain the relation between the star cluster mass,
$M_\mathrm{ecl}$, and its most massive star, $m_\mathrm{max}$, but
the existence of a unique solution can be shown
\citep{pflamm_altenburg2006a}. Using a simple bisection method this relation
can be calculated numerically (Fig. \ref{fig_m-m}, crosses).
The numerical solution can be
very well fitted by the function 
\begin{equation}
  \label{equ_m-m}
  y = a \;x\;   
  (b^n+x^n)^{-1/n}+c  \;\;\;,
\end{equation}
with $y=\log_{10}\left(m_\mathrm{max}/M_\odot\right)$, 
$x=\log_{10}\left(M_\mathrm{ecl}/M_\odot\right)$, $a=2.56$, $b=3.82$,
$c=-0.38$ and $n=9.17$ (Fig. \ref{fig_m-m}, solid line).

This function behaves linearly for  $x\ll b$ 
($y = (a/b) \;x +c$) and is constant for $x\gg b$ (y=a+c). The transition 
occurs at $x\approx b$  whereas a large $n$ describes a rapid transition.
Numerical simulations of star-formation in clusters \citep{bonnell2004a}
also indicate that the mass of the most massive star scales with the system
mass. 

The IGIMF is then calculated by adding all stars in all clusters,
as already noted by \citet{vanbeveren1982a,vanbeveren1983a}, 
\begin{equation}
  \xi_\mathrm{IGIMF}(m) =
  \int_{M_\mathrm{ecl,min}}^{M_\mathrm{ecl,max}(SFR)}
  \xi_{M\mathrm{ecl}}(m)\;\xi_\mathrm{ecl}(M_\mathrm{ecl})\;
  \mathrm{d}M_\mathrm{ecl}\;\;\;.
\end{equation}

\subsection{Number of ionising photons}
\label{ionising_photons}
The total number of ionising photons, $N_{\mathrm{ion},\delta t}(m)$, 
emitted by a star of the mass $m$ within the time 
$\delta t = 10\;\mathrm{Myr}$, is calculated by
integrating the emission rate of ionising photons, $N_\mathrm{ion}(m,t)$,
 over
$\delta t$
\begin{equation}
  \label{eqn_N_ion_delta_t}
  N_{\mathrm{ion},\delta t} = \int_{0}^{\delta t} N_\mathrm{ion}(m,t)\;\;\;
  \mathrm{d}t.
\end{equation}%
\begin{figure}
  \plotone{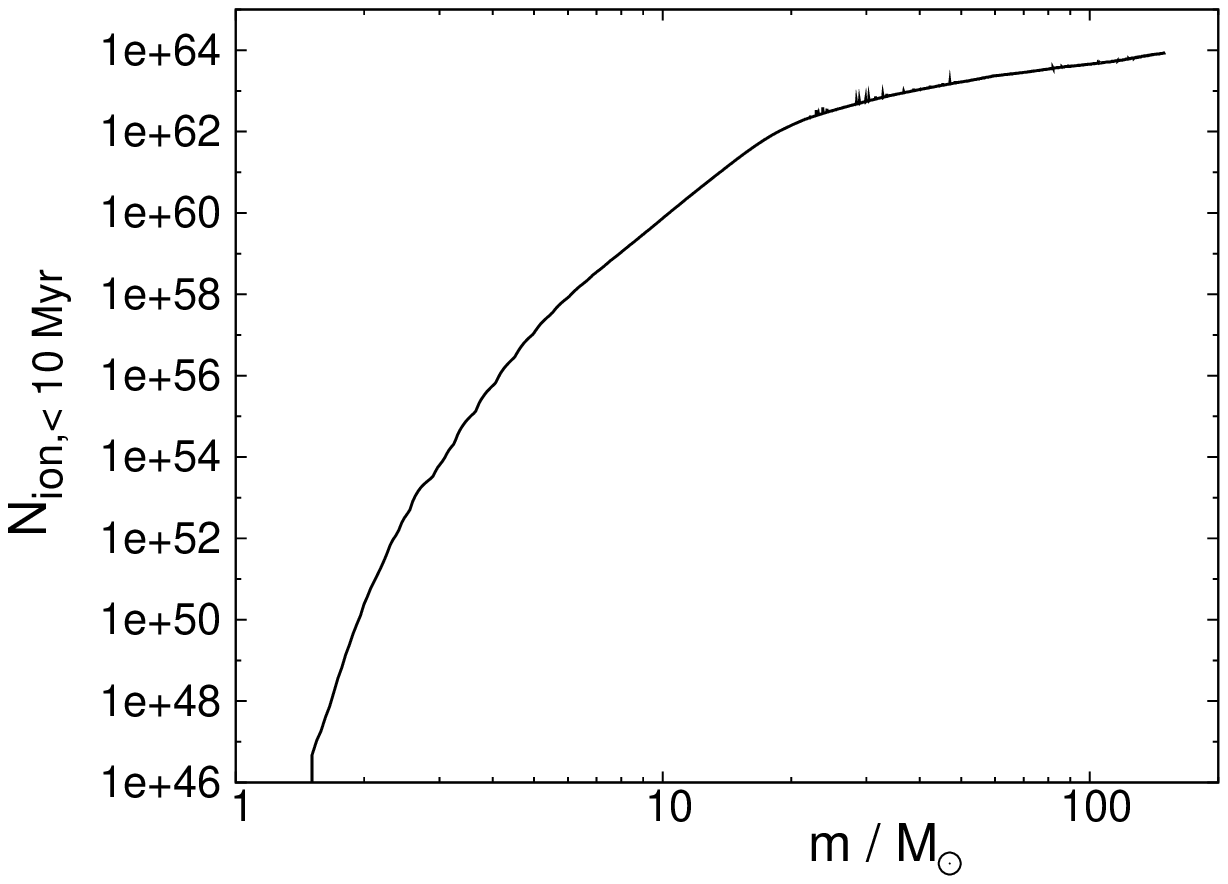}
  \caption{The total number of ionising photons, $N_{\mathrm{ion},\delta t}$, 
    emitted by a star with the mass $m$ in solar masses
    within the time $\delta t$ = 10~Myr 
    at the beginning of the life of the star.}
  \label{fig_N_tot}
\end{figure}%
Stars with life times shorter than $\delta t$ contributes all of their
emitted ionising photons.

In order to derive the emission rates of ionising photons, a grid of
4000 stars linearly spaced between 0.01 and 150~$M_{\odot}$ is evolved
from 0 to 20~Myr in time steps of 0.1~Myr. For stars below 
50~$M_{\odot}$ the stellar evolution package by \cite*{hurley2000a} 
is used. Above that limit until 120~$M_{\odot}$, 
the stellar parameters are interpolated from models by
\citet[without rotation]{meynet2003a}. For the
most massive stars above 120~$M_{\odot}$  formulae fitted to the
massive star models by \citet{schaller1992a}
are used in extrapolation\footnote{An extensive
  description of the fitting formulae can be found in 
\citet{weidner2006a}}. The physical parameters provided by
the models are then used to find appropriate stellar spectra for each
star at every time step from the BaSeL stellar library 
\citep{lejeune1997a,lejeune1998a,westera2002a}.
From these spectra the part of the
luminosity which can ionise hydrogen (all radiation with a wavelength
$\lambda \le$~912~$\rm
\AA$), the ionising luminosity, $L_{\mathrm{ion}}(m,t)$, is deduced for each
star. Finally, the number of the ionising photons, $N_{\mathrm{ion}}(m,t)$, can
be calculated from
\begin{equation}
N_{\mathrm{ion}}(m,t) = L_{\mathrm{ion}}(m,t)\,10^{10.5},
\end{equation}
from \citet{stahler2005a}.
The resulting $N_\mathrm{ion, \delta t}$ (Eq.~\ref{eqn_N_ion_delta_t})
is shown in Fig.~\ref{fig_N_tot}.

\begin{figure}
\plotone{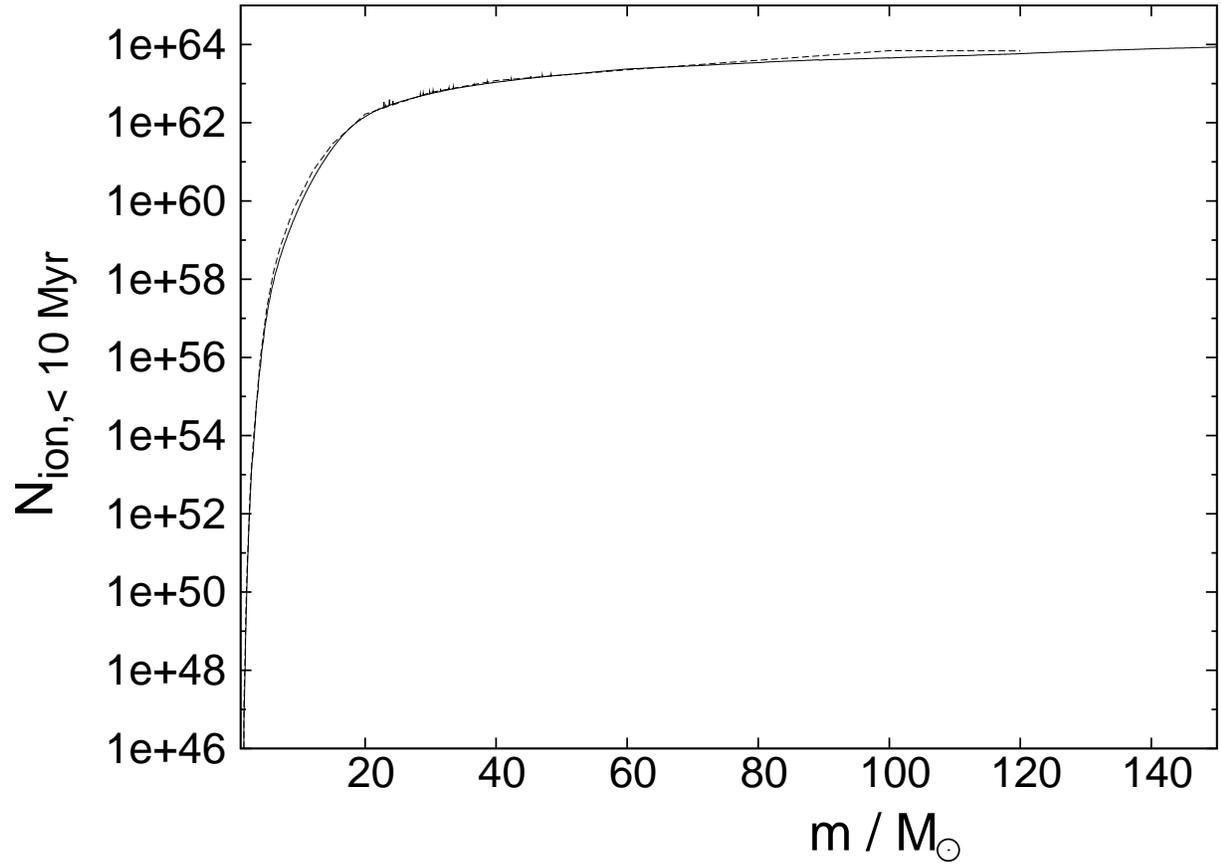}
\caption{Logarithmic number of ionising photons ($N_{\rm ion,< 10 Myr}$) versus
  stellar mass for the here used models ({\it solid line}) and the
  Padova94 tracks ({\it dashed line}).}
\label{fig_Padova94}
\end{figure}

In order to compare the results obtained with the here used set of
stellar evolution models\footnote{A more detailed description of the
  models will be given in Weidner \& Kroupa (2007, in preparation).}
with other models,
the same procedure as described above was applied to the Padova94
tracks for solar metallicity \citep{bressan1993a} which
are preferred by \citet{bruzual2003a}. As \citet{bruzual2003a}
also use the BaSeL library of stellar
spectra it is possible to use our procedure to obtain $N_{\mathrm{ion}}$
but with the different stellar models. 
$N_{\mathrm{ion}}$ is plotted in Fig.~\ref{fig_Padova94} in 
dependence of stellar mass using the Padova94 tracks
({\it dashed line}) and using our models ({\it solid line}). Both
compare very well.
\subsection{Models}
\label{sec_models}
Although the IMF of young star clusters seems to be universal and
  widely independent of their environment \citep{kroupa2001a,kroupa2002a}, 
  the universality of the ECMF slope is not an established result as
  summarised in \citet{weidner2005a}. To be consistent with  our
  previous work, the four IGIMF models here are the same as defined in 
  \citet{weidner2005a}. Both the IMF and the ECMF
  are multi-power laws, $\xi(m) \propto m^{-\alpha_\mathrm{i}}$ 
  and $\xi_\mathrm{ecl}(M_\mathrm{ecl})\propto
  M_\mathrm{ecl}^{-\beta_\mathrm{i}}$. 
  All slopes and mass limits of all four IGIMF-scenarios 
  are summarised in Tab. \ref{tab_igimf_scenarios}.

\begin{deluxetable}{ccccc}
\tabletypesize{\normalsize}
\tablewidth{0pt}
\tablecaption{Slopes and mass limits of the IMF and ECMF for four IGIMF-models
\label{tab_igimf_scenarios}}
\tablehead{\colhead{Parameter}&\colhead{Standard}&\colhead{Minimal-1}
&\colhead{Minimal-2}&\colhead{Maximal}}
\startdata
$m_1$/M$_\odot$       &0.08&0.08&0.08& 0.08\\
$\alpha_1$  &1.30&1.30&1.30& 1.30\\
$m_2$/M$_\odot$       &0.5&0.5&0.5&0.5 \\
$\alpha_2$  &2.35&2.35&2.35& 2.35\\
$m_3$/M$_\odot$       &1.0&1.0&1.0&1.0 \\
$\alpha_3$  &2.35&2.35&2.35&2.7 \\
$M_1$/M$_\odot$       &5&5&\nodata& 5\\
$\beta_1$   &2.35&1.0&\nodata&2.35\\
$M_2$/M$_\odot$       &50&50&50&50\\
$\beta_2$   &2.35&2.0&2.0&2.35\\
\enddata
\tablecomments{
The IMF slopes $\alpha_i$ refer to a power law, $\xi(m)\propto
m^{-\alpha_\mathrm{i}}$,
on a mass interval between $m_\mathrm{i}$ and $m_\mathrm{i+1}$
where $m_\mathrm{i+1}\le m_\mathrm{max*}$. 
The ECMF slopes $\beta_i$ refer to a power-low,
$\xi_\mathrm{ecl}\propto M_\mathrm{ecl}^{-\beta_\mathrm{i}}$,
on a mass interval between $M_\mathrm{i}$ and $M_\mathrm{i+1}$
where
$M_\mathrm{i+1} = M_\mathrm{ecl,max}$(SFR)
(Eq. \ref{eqn_m_ecl_max_sfr}).}
\end{deluxetable}

The {\it Standard-IGIMF} is based  on a canonical IMF
(Eq.~\ref{equ_imf_slopes}) and an ECMF with a power $\beta=2.35$
between 5~M$_\odot$ and $M_\mathrm{ecl,max}$(SFR). A slope
of 2.35 provides the best agreement between the theoretical and the observed 
relation between the actual SFR and the mass of the heaviest young star cluster
\citep{weidner2004b}. Furthermore, the theoretical formation time scale 
of star clusters has a constant value of 10~Myr independent of the cluster mass
if a slope of 2.35 is chosen.

In order to explore the dependence of the IGIMF on the ECMF
we test three additional scenarios:

The {\it Minimal-1 Scenario} has a  slope of $\beta_1=1$ 
for small cluster masses  ($5 \le M_\mathrm{ecl}/\mathrm{M}_\odot \le 50$) 
and $\beta_2 = 2$ for higher masses 
($50 \le M_\mathrm{ecl}/\mathrm{M}_\odot \le M_\mathrm{ecl,max}$)
and a canonical IMF (Eq.~\ref{equ_imf_slopes}).

The {\it Minimal-2 Scenario} has a truncated ECMF with $\beta = 2$ 
($50 \le M_\mathrm{ecl}/\mathrm{M}_\odot \le M_\mathrm{ecl,max}$), 
i.e. no star clusters below 50~M$_\odot$ are formed,
and a canonical IMF (Eq.~\ref{equ_imf_slopes}). 

The {\it Maximal Scenario} consists of an IMF
(Eq.~\ref{equ_imf_slopes}) with a steeper high mass star slope
of $\alpha_3 = 2.7$ and an ECMF with $\beta=2.35$
between 5~$M_\odot$ and $M_\mathrm{ecl,max}$(SFR).
  
Both minimal scenarios produce a larger number of 
massive star clusters and therefore more massive stars 
than the Standard Scenario. 
The Maximal Scenario produces fewer massive stars than the Standard Scenario.
Thus, for a given SFR both minimal scenarios produce a higher 
H$\alpha$ luminosity and the Maximal Scenario a lower H$\alpha$ luminosity
than the Standard Scenario.

In order to explore the change of the
classical linear $L_{\mathrm{H}\alpha}$-SFR relation due to the effect 
of the IGIMF we compare the IGIMF models with two scenarios in which
the IGIMF is invariant, i.e. independent of the SFR.
These invariant scenarios as well as the linear scenarios 
by \citet{kennicutt1994a} provide theoretical H$\alpha$ luminosities 
based on the assumption that the galactic-wide distribution function
of newly formed
stars corresponds  to the IMF observed in young clusters. Thus, the
slope of the IMF is independent of the total SFR and the IGIMF
is identical to the IMF in young star clusters. 

In our {\it first invariant model} the IMF is chosen to have the canonical form
defined by Eq. \ref{equ_imf_slopes} and uses the same stellar properties
calculated in Sec. \ref{ionising_photons} 
where $m_\mathrm{max}=m_\mathrm{max*}=$~150~M$_\odot$.

Our {\it second invariant model} consists of a Salpeter IMF with a slope of 2.35
between 0.1 and 100~M$_\odot$ and the same stellar properties
calculated in Sec. \ref{ionising_photons}. This IMF is the basis of
the widely used {\it linear model} derived by 
\citet{kennicutt1994a}, 
\begin{equation}
  \label{eqn_kennicutt}
  \mathrm{SFR}(\mathrm{total}) = \frac{L_{\mathrm{H}\alpha}}
  {1.26\cdot10^{41}\;\mathrm{erg}\;\mathrm{s}^{-1}}\;\mathrm{M}_\odot\;
  \mathrm{yr}^{-1}\;\;\;.
\end{equation}

\subsection{Results}
\label{sec_results}

\begin{figure}
  \plotone{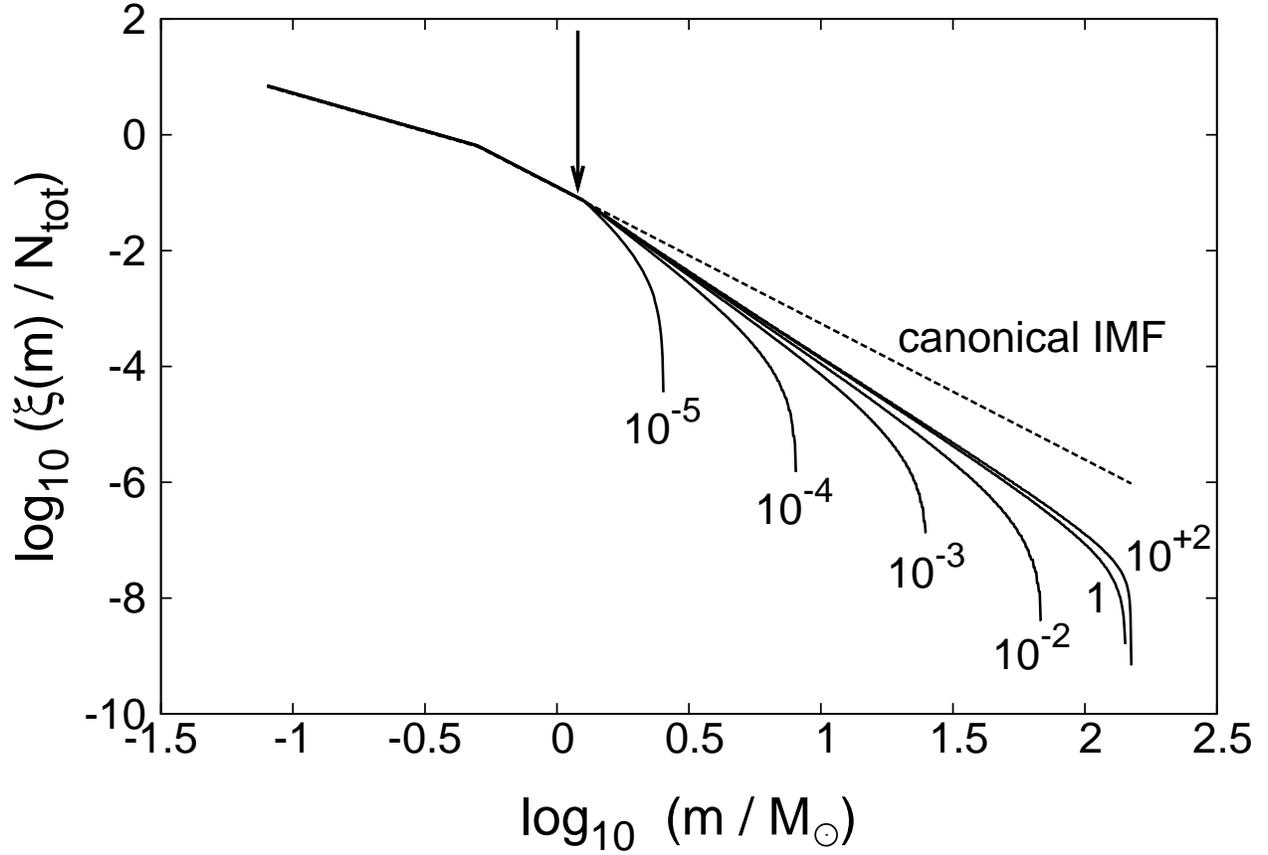}
  \caption{Normalised IGIMF in the Standard Scenario 
    for different SFRs (10$^{-5}$, 10$^{-4}$,
    10$^{-3}$, 10$^{-2}$, 1, 100~M$_\odot$~yr$^{-1}$), 
    and the canonical IMF (upper-most line). The arrow indicates
    the mass at which all IGIMFs start to deviate from the underlying
    canonical IMF ($m_\mathrm{max}(M_\mathrm{ecl,min}) = 1.25~\mathrm{M}_\odot$).
    Note that the IGIMFs are not exactly identical for 
    $m\le 1.25~\mathrm{M}_\odot$ due to the normalisation by the total 
    mass of stars in each case.
 }
  \label{fig_igimf}
\end{figure}
The linearity of the widely used relation, Eq.~\ref{eqn_kennicutt},
implies that if the total SFR is reduced by a certain factor then the
produced H$\alpha$ luminosity is reduced by the same factor, because the
portion of ionising massive stars is constant if the IGIMF is identical
to a universal IMF. But our concept of the IGIMF is a combination of two
effects: firstly, with decreasing SFR the upper limit of the ECMF falls
(Eq.~\ref{eqn_m_ecl_max_sfr}) and secondly the
fraction of ionising massive stars is higher in heavy star clusters than in
light star clusters due to the relation between the mass of a star 
cluster and its most massive star (Eq.~\ref{equ_m-m}). 
Therefore, the galactic-wide fraction of massive
stars is expected to decrease with decreasing total SFR.

The resulting normalized Standard-IGIMF, i.e. the Standard-IGIMF 
divided by the total number of stars, is plotted in Fig. \ref{fig_igimf} 
for different SFRs. As expected, all IGIMFs are steeper than 
the canonical IMF in the high-mass regime, and the mass of the most-massive
star of the entire galaxy decreases with decreasing SFR. 
All IGIMFs start to deviate (Fig.~\ref{fig_igimf}, arrow) 
from each other above a mass threshold
which is the maximum stellar mass of the smallest star cluster
($m_\mathrm{max}(M_\mathrm{ecl,min}=5\;\mathrm{M}_\odot)=~$1.25~M$_\odot$).
Stars below this threshold are formed in all star clusters. 
Note also, that the IGIMF has a lower upper mass limit than the 
canonical IMF.

\begin{figure}
  \plotone{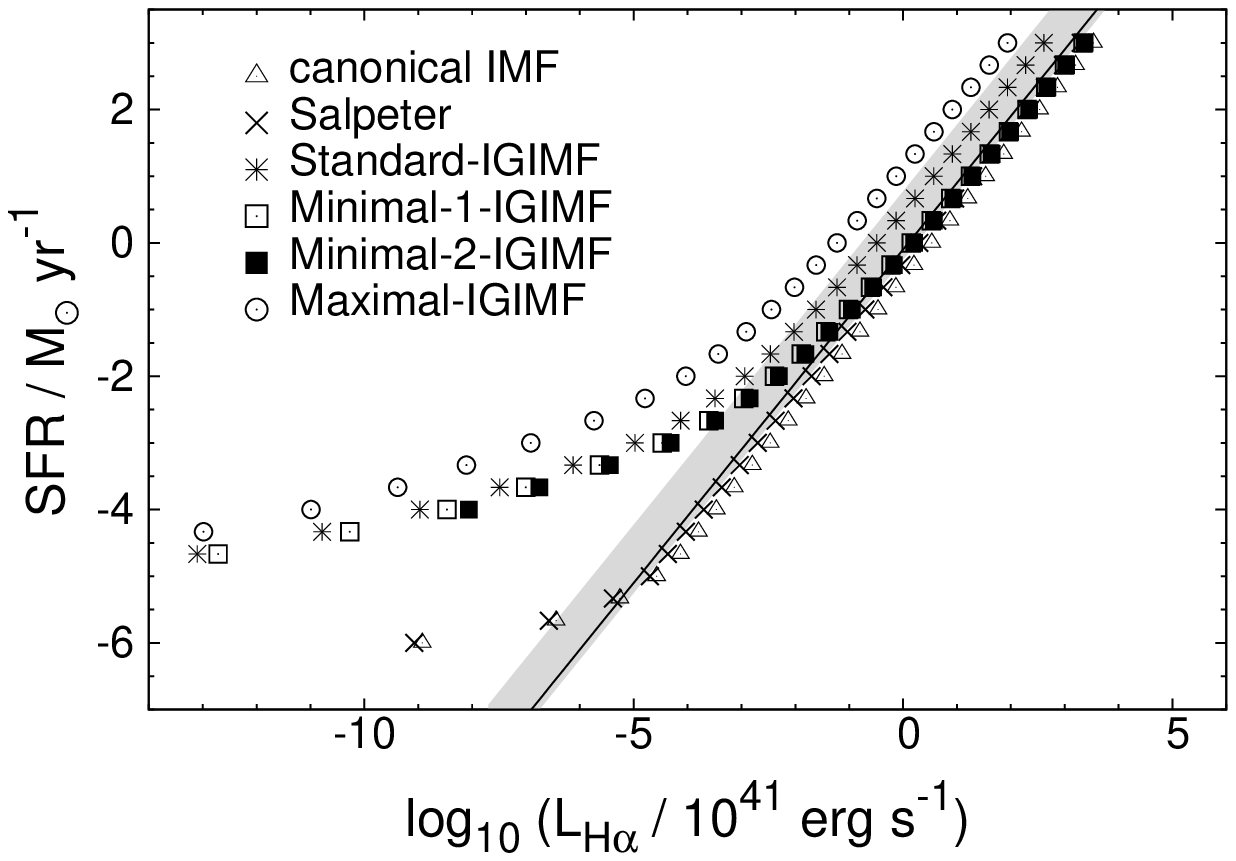}
  \caption{The calculated relation between the SFR and
    the resulting H$\alpha$-luminosity for four different IGIMF-models
    and two invariant models in which the IGIMF is assumed 
    to be the canonical  IMF and the Salpeter IMF. 
    The solid line represents the widely used
    linear relation, Eq. \ref{eqn_kennicutt},
    by \citet{kennicutt1994a}. The gray shaded area marks the full set of
    linear $L_{\mathrm{H}\alpha}$-SFR relations \citep{kennicutt1994a}
    combining different stellar evolution models and different IMFs
    (for details see text).}
  \label{fig_Lalpha_SFR}
\end{figure}

Because the functional form of the IGIMF is a function of the total SFR
the produced H$\alpha$ luminosity is expected to not depend linearly of
the total SFR.
The resulting relation between the SFR and the produced H$\alpha$
luminosity is plotted in Fig. \ref{fig_Lalpha_SFR}: This includes all
four IGIMF-models and the two invariant models described
above. Additionally, the widely used linear 
$L_{\mathrm{H}\alpha}$-SFR 
relation, Eq. \ref{eqn_kennicutt} \citep[solid line]{kennicutt1994a},
and the complete sample of linear models based on different stellar
evolution models and different IMFs \citep[gray shaded
area]{kennicutt1994a} are plotted, too. 

The IGIMF models are nearly linear above an
H$\alpha$ luminosity of about 10$^{40}$~erg~s$^{-1}$, which is the lower
value of the normal disk galaxies explored by 
\citet{kennicutt1983a}, whereas the relations become much flatter below
10$^{40}$~erg~s$^{-1}$. As expected the IGIMF-models lie 
above the linear relations. The order of magnitude of the deviation
of the IGIMF-$L_{\mathrm{H}\alpha}$-SFR relation from the linear 
$L_{\mathrm{H}\alpha}$-SFR relation is independent 
of the explicit choice of the IGIMF model.

For high $H\alpha$ luminosities our 
invariant models are strongly linear like the Kennicutt relations.
Below a SFR of about 10$^{-5}$~M$_\odot$~yr$^{-1}$ they become shallower.
In our invariant model the IGIMF is replaced by the canonical IMF.
But the IMF is normalised such that the total mass contained in it
is still given by 
\begin{equation}
  \mathrm{SFR}\;\delta t  =
  \int_{m_\mathrm{min}}^{m_\mathrm{max}}m\;\xi(m)\;
  \mathrm{dm}\;\;\;,
\end{equation}
where $m_\mathrm{max}$ is the minimum of the physical upper mass limit,
$m_\mathrm{max*}$ and the total mass content, $\mathrm{SFR}\;\delta t$
(a star cannot be more massive than the total mass out of which it is
formed). Note that $\xi(m) = \xi_\mathrm{Mecl}(m)$
(Eq. \ref{equ_imf_slopes})
 for  $m_\mathrm{max} = m_\mathrm{max*}$.
If the mass available for star formation within the epoch
$\delta t$ is less than the physical upper mass limit $m_\mathrm{max*}$
then the upper limit of the IMF is given by the amount of available
mass. Thus, at a SFR of
\begin{equation}
  \mathrm{SFR} = \frac{m_\mathrm{max*}}{\delta t} = 
  \frac{150\;\mathrm{M}_\odot}{10\;\mathrm{Myr}} 
  = 10^{-4.82}\;\frac{\mathrm{M}_\odot}{\mathrm{yr}}
\end{equation}
our $L_\mathrm{H\alpha}$-SFR relations diverge from linearity. The simple
physical boundary condition that stars more massive the $SFR\;\delta t$
cannot form leads to non-linear behaviour of the $L_\mathrm{H\alpha}$-SFR
relation.

For dwarf irregular galaxies having H$\alpha$ luminosities between
10$^{-36}$ and 10$^{-39}$~erg~s$^{-1}$ the underestimation of the SFRs
vary by a factor of at least 20 to 160 for the Minimal Scenario.
In the  Maximal Scenario the underestimation of SFRs vary between a factor
of 800  and 5000. The SFRs of normal disk galaxies with an H$\alpha$ luminosity
of about 10$^{40}$--10$^{42}$~erg~s$^{-1}$ may be underestimated
by a factor of 4 up to 50.

\subsection{Fitting functions}
\begin{figure}
  \plotone{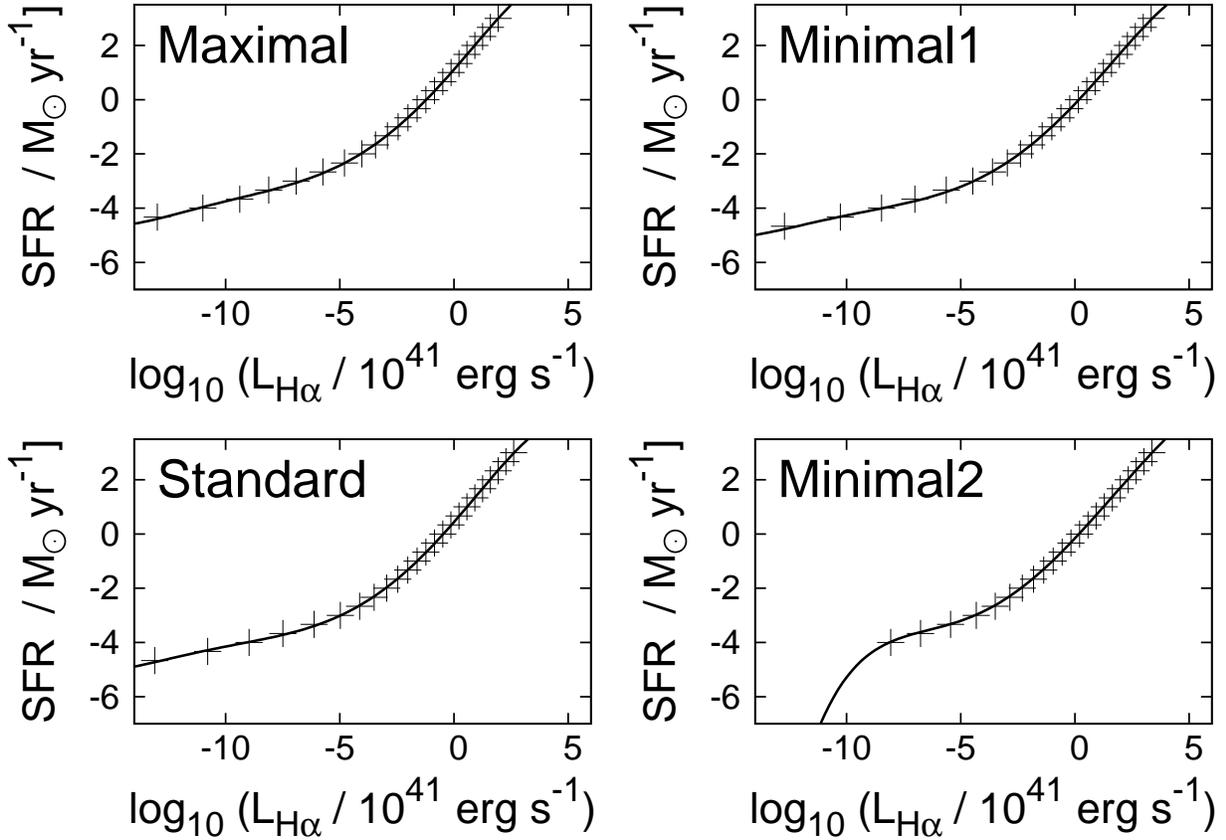}
  \caption{Fitted plot of the four IGIMF $L_{\mathrm{H}\alpha}$-SFR relations
    (crosses) using the fit function Eq. \ref{equ_fitting_function} 
  and the parameters of Tab. \ref{tab_parameter} (solid lines).}
  \label{fig_sfr_fit}
\end{figure}

For convenient use of the obtained $L_{\mathrm{H}\alpha}$-SFR relation 
we provide fitting functions for each IGIMF-model.
The general form of the fitting function  is a polynomial of fifth-order,
\begin{equation}
  \label{equ_fitting_function}
  y = a_5x^5+a_4x^4+a_3x^3+a_2x^2+a_1x+a_0\;\;\;.
\end{equation}
This function is used to fit the data-points of 
each IGIMF-model (Fig. \ref{fig_Lalpha_SFR}).
The argument and the function value of the fitting-function
are
\[
x = \log_{10}\left(L_{\mathrm{H}\alpha} 
  \left[ 10^{41} \mathrm{erg} \;\mathrm{s}^{-1}\right]\right) \;\;\;, 
\]
\begin{equation}
  f(x) = \log_{10}\left( \mathrm{SFR} 
    \left[ \mathrm{M}_\odot \;\mathrm{yr}^{-1}\right]\right)\;\;\;.
\end{equation}
The obtained values of the fitting parameters are listed in 
Tab.~\ref{tab_parameter} for each model and the fitted models are plotted in 
Fig.~\ref{fig_sfr_fit}.
\begin{deluxetable}{lccccccc}
  \tabletypesize{\normalsize}
  \tablewidth{0pt}
  \tablecaption{List of the fitting parameters of the 
    $L_{\mathrm{H}\alpha}$-SFR relation\label{tab_parameter}}
  \tablehead{\colhead{}&\colhead{}&\colhead{}&  
    \colhead{}&\colhead{}&\colhead{}&\colhead{} 
    &\colhead{$L_{\mathrm{H}\alpha}$-range}\\
    \colhead{Model}&\colhead{$a_5$}&\colhead{$a_4$}&  
    \colhead{$a_3$}&\colhead{$a_2$}&\colhead{$a_1$}&\colhead{$a_0$} 
    &\colhead{(erg s$^{-1}$) }
  }
  \startdata
    Standard&-2.67e-05&-9.30e-04&-8.47e-03&+3.21e-02&+9.64e-01&+4.38e-01&
    $1.7\cdot 10^{25}$ -- $4.1\cdot 10^{43}$\\
    Maximal&-2.94e-05&-9.97e-04&-9.13e-03&+2.67e-02&+9.63e-01&+1.11e-00&
    $7.4\cdot 10^{25}$ -- $1.2\cdot 10^{43}$\\
    Minimal-1&-2.32e-05&-7.51e-04&-5.70e-03&+4.23e-02&+8.85e-01&-1.58e-01&
    $3.1\cdot 10^{25}$ -- $2.2\cdot 10^{44}$\\
    Minimal-2&3.84e-05&-2.43e-04&-6.58e-03&+3.50e-02&+8.94e-01&-1.91e-01&
    $8.6\cdot 10^{32}$ -- $2.3\cdot 10^{44}$\\
  \enddata
  \tablecomments{
    Fit-parameters of the fitting-function 
    (Eq. \ref{equ_fitting_function}) for the four different IGIMF-models:
    Standard, Maximal, Minimal-1 and Minimal-2 and the luminosity range of
    data points used for the fit.} 
\end{deluxetable}
The linear parts of our invariant models can be described by
\begin{eqnarray}
  \mathrm{SFR} = \frac{L_{\mathrm{H}\alpha}}
  {1.89\cdot 10^{41}\;\mathrm{erg}\;\mathrm{s}^{-1}}\;
  \mathrm{M}_\odot\;\mathrm{yr}^{-1}\;\;,\nonumber\\
  \mathrm{for}\;\;\;L_{\mathrm{H}\alpha} \ge 2.8\cdot 10^{36}\;\mathrm{erg}\;\mathrm{s}^{-1}
\end{eqnarray}
in the case of the canonical IMF and
\begin{eqnarray}
  \mathrm{SFR} = \frac{L_{\mathrm{H}\alpha}}
  {3.3\cdot 10^{41}\;\mathrm{erg}\;\mathrm{s}^{-1}}\;
  \mathrm{M}_\odot\;\mathrm{yr}^{-1}\;\;,\nonumber\\
  \mathrm{for}\;\;\;L_{\mathrm{H}\alpha} \ge 4.9\cdot 10^{36}\;\mathrm{erg}\;\mathrm{s}^{-1}
\end{eqnarray}
in the case of the Salpeter- IMF.
\section{SFRs of dIrr}
\label{sec_sfr_dIrr}
To study the effect of the IGIMF in real systems 
we now calculate the SFRs of the Sculptur dwarf irregular galaxies. 
The H$\alpha$ luminosities are taken from 
\citet{skillman2003a} and are already corrected for galactic extinction 
but not for  N {\sc ii} contamination as these dwarf galaxies have typically 
very low nitrogen abundances. The SFRs are obtained using the fitting function
Eq. (\ref{equ_fitting_function}).
The SFRs in \citet{skillman2003a} are based on the classical relation
Eq. \ref{eqn_kennicutt}.

\begin{figure}
  \plotone{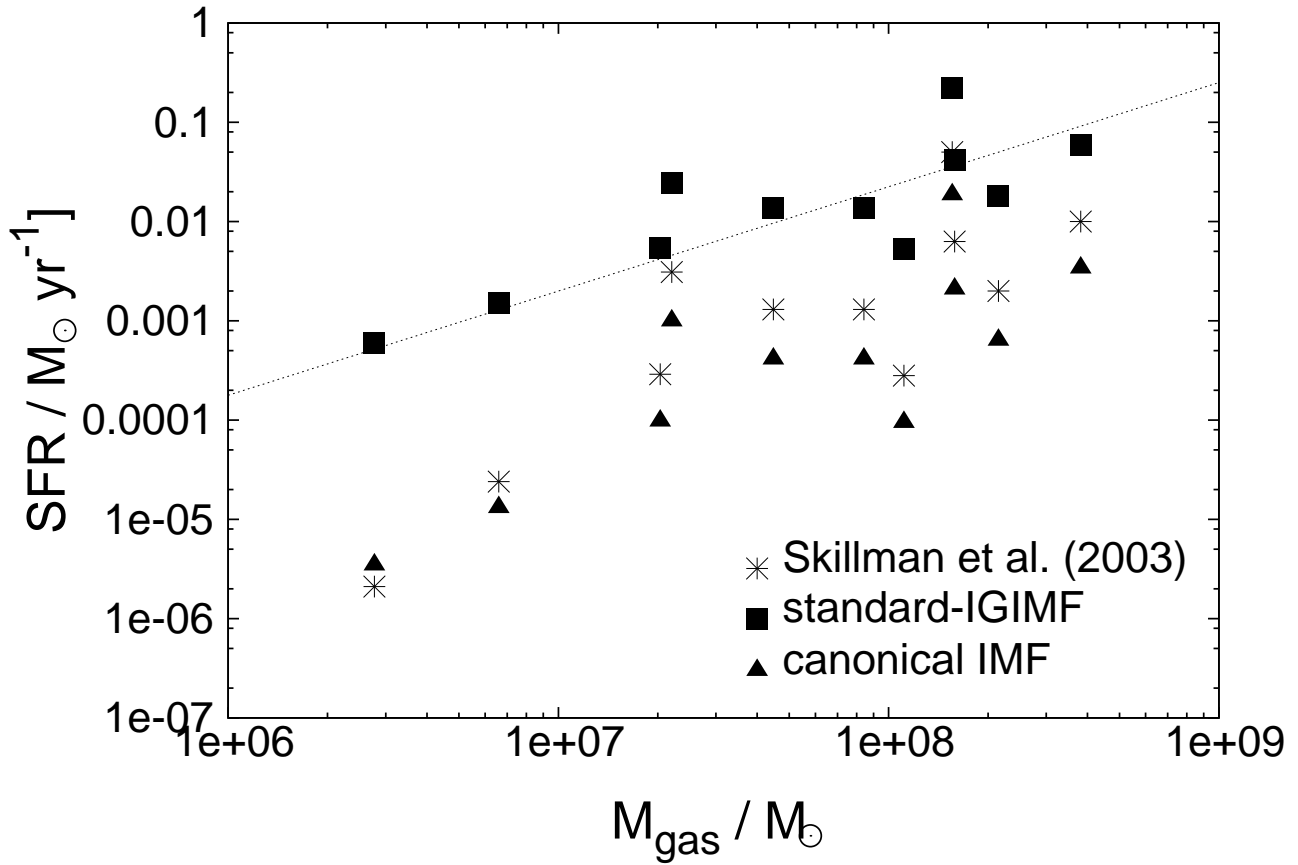}
  \caption{SFRs in Sculptor dwarf irregular galaxies in dependence of the
    total galaxy gas mass. The galaxy gas masses and the SFRs calculated using
    the Kennicutt-relation are taken from
    \citet[Tab. 3]{skillman2003a}. These SFRs are compared with the ones
    obtained using the standard-IGIMF relation (SFR$_\mathrm{STD}$ values in 
    Tab. \ref{tab_parameter}) and with our invariant model where the IGIMF
    is replaced by a canonical IMF.
    The solid line marks the bivariate regression to the data calculated
    with the Standard-IGIMF described 
    by Eq.~\ref{dIrr_SFR_fit}.}
\label{fig_sculptor-dIrr}
\end{figure}

Observed H$\alpha$ luminosities, linear and IGIMF-SFRs and H {\sc i} masses,
$M_\mathrm{H I}$ ,
for eleven Sculptor dwarf irregular galaxies are listed in Tab. 
\ref{tab_sculptor-dIrr}.
The relation between the total galaxy gas mass, $M_\mathrm{gas}$, 
and the SFR can be seen in
Fig. \ref{fig_sculptor-dIrr}. To account for Helium 
the total galaxy gas masses are calculated from
\begin{equation}
  M_\mathrm{gas} = 1.32 \; M_{\mathrm{H I}}\;\;\;, 
\end{equation}
following \citet{skillman2003a}.

The conventional data based on the
Kennicutt relation (asterisks in Fig.~\ref{fig_sculptor-dIrr}) can
be divided roughly into two parts. Above a galaxy gas mass of
about $2\cdot 10^{7}$~M$_\odot$ the SFRs are uniformly distributed 
around a value
of about $7\cdot 10^{-3}$~M$_\odot$~yr$^{-1}$ with a large scatter ranging from 
$2.8\cdot 10^{-4}$ to $5.0\cdot 10^{-2}$~M$_\odot$~yr$^{-1}$. 
Below a galaxy gas mass of
about $2\cdot 10^{7}$~M$_\odot$ there is a steep decline of the SFR with
decreasing galaxy gas mass.
This relation is changed by the standard-IGIMF-effect (filled squares)
in three ways: Firstly, 
all SFRs
are increased. This increase is larger for low SFRs and therefore 
mainly for small galaxy gas masses. Secondly, the scatter of SFRs above     
a total galaxy gas mass of about $2\cdot 10^{7}$~M$_\odot$ reduces by about one
order of magnitude. Thirdly, the decrease of the SFR with decreasing 
total galaxy gas mass becomes less steep. 

A bivariate linear regression  between the SFRs derived in the 
Standard-IGIMF model and the total galaxy gas mass 
(Fig.~\ref{fig_sculptor-dIrr}, solid line) leads to the relation 
\begin{equation}
  \label{dIrr_SFR_fit}
  \log_{10}\frac{SFR}{\mathrm{M}_\odot\;\mathrm{yr}^{-1}}
  = 1.05 \cdot \log_{10}\frac{M_\mathrm{gas}}{\mathrm{M}_\odot}  -10.05\;\;\;.
\end{equation} 
This means that the SFRs of the Sculptor dwarf irregular galaxies
depends linearly on the total gas mass:
\begin{equation}
  SFR = 8.9\cdot 10^{-11}\;\frac{\mathrm{M}_\odot}{\mathrm{yr}^{-1}}\;\;
  \cdot M_\mathrm{gas}^{1.05}\;\;\;.
\end{equation}

\begin{figure}
  \plotone{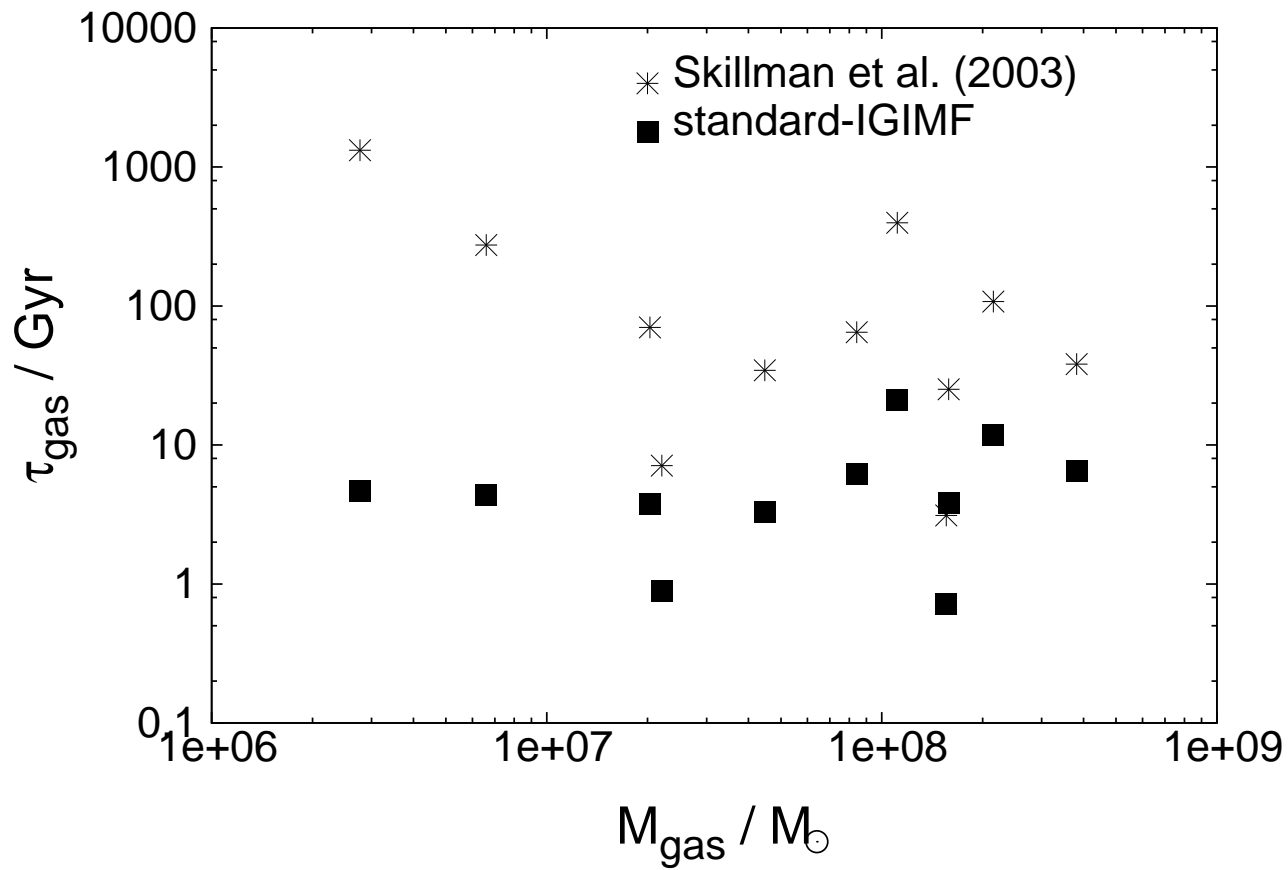}
  \caption{Gas depletion times  in Sculptor dwarf irregular galaxies 
    in dependence of the total galaxy gas mass. 
    The galaxy gas masses and the SFRs calculated using
    the $L_\mathrm{H\alpha}$-SFR relation by Kennicutt are taken from
    \citet[Tab. 3]{skillman2003a}. These SFRs are compared with the ones
    obtained using the Standard-IGIMF $L_\mathrm{H\alpha}$-SFR relation 
    (Tab. \ref{tab_parameter}).}
  \label{fig_sculptor-dIrr_depletion}
\end{figure}

The corresponding gas depletion time scales,
\begin{equation}
  \tau_\mathrm{gas} = M_\mathrm{gas}/ SFR\;\;\;,
\end{equation}
are shown in Fig. \ref{fig_sculptor-dIrr_depletion}. 
While the gas depletion times vary by a factor of 400 in the linear
(Kennicutt) picture, the IGIMF leads to a shrinkage of the range 
down to only a factor of 20.
Additionally, the strong increase of the gas depletion time with lower galaxy
gas mass disappears and the gas depletion time becomes constant at a few Gyr.
This is not surprising, as a strictly linear dependence of the SFR
on the total gas mass,
\begin{equation}
  SFR = AM_\mathrm{gas}\;\;\;,
\end{equation}
directly implies a constant gas depletion time scale,
\begin{equation}
  \tau_\mathrm{gas} = M_\mathrm{gas}/SFR = A^{-1}\;\;\;.
\end{equation}

\begin{deluxetable}{cccccccc}
  \tabletypesize{\normalsize}
  \tablewidth{0pt}
  \tablecaption{SFRs of Sculptor dIrr Galaxies
  \label{tab_sculptor-dIrr}}
  \tablehead{\colhead{}&\colhead{$M_\mathrm{H {\sc I}}$}&
    \colhead{$\log_{10}(L_{\mathrm{H}\alpha})$}&\colhead{SFR$_\mathrm{Skill}$}& 
    \colhead{SFR$_\mathrm{STD}$}&\colhead{SFR$_\mathrm{MIN1}$}&
    \colhead{SFR$_\mathrm{MIN2}$}&\colhead{SFR$_\mathrm{MAX}$}\\
  \colhead{Galaxy}&\colhead{(10$^6$~M$_\odot$)}&
    \colhead{(erg~s$^{-1}$)}&\colhead{(M$_\odot$~yr$^{-1}$)}& 
    \colhead{(M$_\odot$~yr$^{-1}$)}&\colhead{(M$_\odot$~yr$^{-1}$)}&
    \colhead{(M$_\odot$~yr$^{-1}$)}&\colhead{(M$_\odot$~yr$^{-1}$)}}
  \startdata
  ESO 347-G17  & 120&38.9 & $6.3\cdot\;10^{-3}$&$4.1\cdot\;10^{-2}$&
  $1.6\cdot\;10^{-2}$&$1.4\cdot\;10^{-2}$&$1.9\cdot\;10^{-1}$\\

  ESO 471-G06  & 163&38.4 & $2.0\cdot\;10^{-3}$&$1.8\cdot\;10^{-2}$&
  $7.9\cdot\;10^{-3}$&$6.6\cdot\;10^{-3}$&$8.0\cdot\;10^{-2}$\\

  ESO 348-G09  &84.3 &37.5 & $2.8\cdot\;10^{-4}$& $5.0\cdot\;10^{-3}$&
  $2.6\cdot\;10^{-3}$&$2.2\cdot\;10^{-3}$&$2.1\cdot\;10^{-2}$\\
  
  SC 18        & 5.0&36.5 & $2.4\cdot\;10^{-5}$& $1.5\cdot\;10^{-3}$&
  $9.3\cdot\;10^{-4}$& $8.4\cdot\;10^{-4}$&$6.2\cdot\;10^{-3}$\\
  
  NGC59        &16.7&38.6 & $3.1\cdot\;10^{-3}$& $2.5\cdot\;10^{-2}$&
  $1.0\cdot\;10^{-2}$& $8.8\cdot\;10^{-3}$&$1.1\cdot\;10^{-1}$\\
  
  ESO 473-G24  &63.8&38.2 & $1.3\cdot\;10^{-3}$&$1.3\cdot\;10^{-2}$&
  $6.0\cdot\;10^{-3}$&$5.0\cdot\;10^{-3}$&$5.8\cdot\;10^{-2}$\\
  
    SC 24        &2.1&35.4 & $2.1\cdot\;10^{-6}$&$5.8\cdot\;10^{-4}$&
    $4.0\cdot\;10^{-4}$&$4.0\cdot\;10^{-4}$&$2.2\cdot\;10^{-3}$\\

    DDO226  & 33.9&38.2 & $1.3\cdot\;10^{-3}$&$1.3\cdot\;10^{-2}$&
    $6.0\cdot\;10^{-3}$&$5.0\cdot\;10^{-3}$&$5.8\cdot\;10^{-2}$\\

    DDO6  & 15.4&37.6 & $2.9\cdot\;10^{-4}$&$5.7\cdot\;10^{-3}$&
    $2.9\cdot\;10^{-3}$&$2.4\cdot\;10^{-3}$&$2.4\cdot\;10^{-2}$\\

    NGC 625      & 118&39.8 & $5.0\cdot\;10^{-2}$& $2.2\cdot\;10^{-1}$&
    $7.1\cdot\;10^{-2}$& $6.3\cdot\;10^{-2}$& $1.0\cdot\;10^{-0}$\\    
    
    ESO 245-G05   & 289 & 39.1& $1.0\cdot\;10^{-2}$& $5.9\cdot\;10^{-2}$&
    $2.2\cdot\;10^{-2}$& $1.9\cdot\;10^{-2}$& $2.7\cdot\;10^{-1}$\\
  \enddata
  \tablecomments{
  Integrated H$\alpha$ luminosities and 
  determined SFRs by \citet{skillman2003a} using the Kennicutt-relation
  (Eq. \protect\ref{eqn_kennicutt}). The SFRs of the dIrrs for the 
  IGIMF-models are calculated using our fitting functions.}
\end{deluxetable}

\section{H$\alpha$-invisible SF}
\label{sec_invisible_sf}
The determination of the SFR of a galaxy is based on the measurement
of a total H$\alpha$ luminosity of the galaxy and a reliable theoretical 
relation between the current SFR and the produced H$\alpha$ radiation.
The H$\alpha$ emission is confined within distinct H {\sc ii}-regions. These
regions have to be identified manually after H$\alpha$ imaging.
The emission of these H$\alpha$ regions have to exceed some detection limit
to be distinguishable from background radiation and noise 
and to be detectable.

\begin{figure}
  \plotone{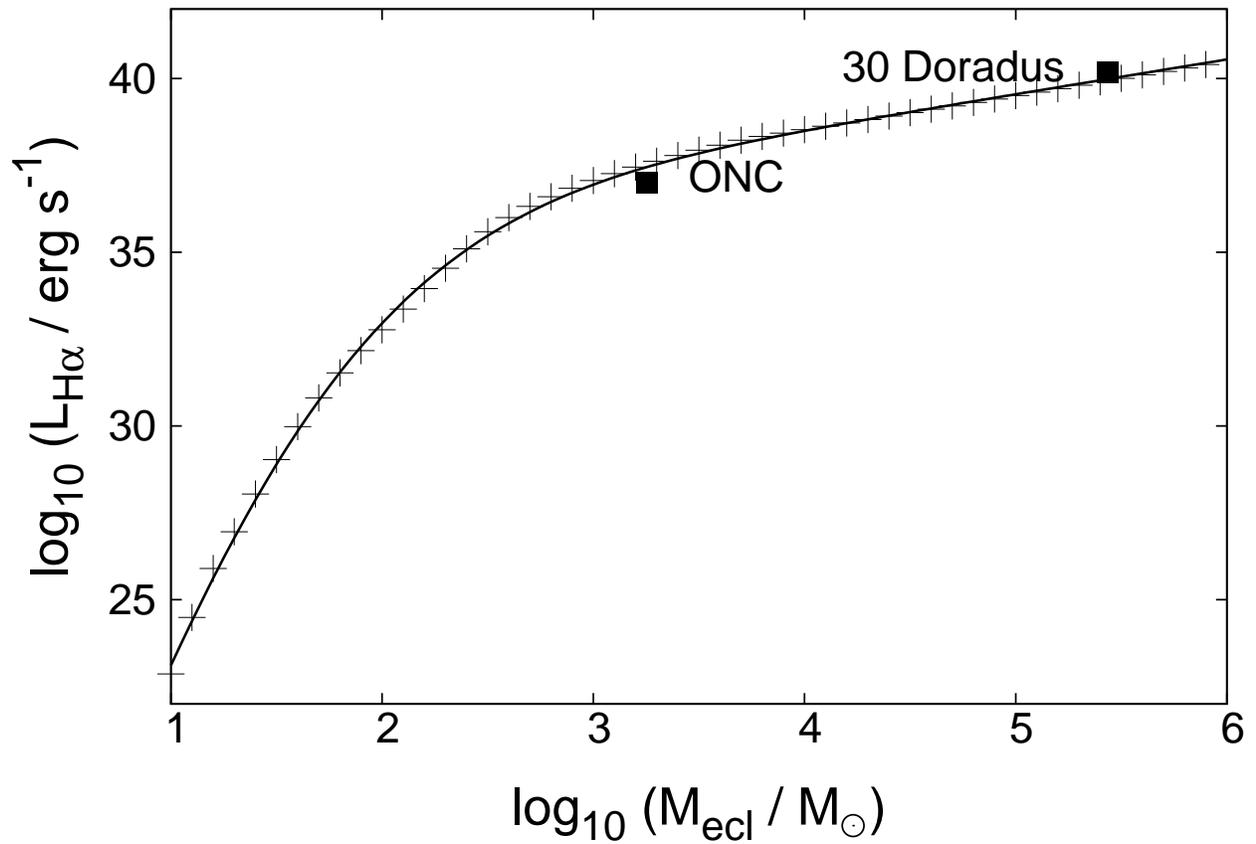}
  \caption{Mass-luminosity relation between the stellar mass of the cluster
    and the produced total H$\alpha$ emission. The solid line shows the fit
    by Eq. \ref{equ_mass_luminosity}. The observed 
    position of the Orion Nebula
    cluster and 30 Doradus are indicated for comparison. Note the steep 
    decline of $L_\mathrm{H\alpha}$ for 
    $M_\mathrm{ecl} \lesssim$~1000~M$_\odot$ is a result of the 
    $m_\mathrm{max}(M_\mathrm{ecl})$ relation (Fig.~\ref{fig_m-m}).}
  \label{fig_mass_luminosity}
\end{figure}

Assuming that SF
takes place in star clusters distributed according to an ECMF with star
cluster masses down to a few solar masses then the least luminous observed 
H {\sc ii}~region is not the smallest star cluster but the cluster
producing the lowest observable H$\alpha$~radiation. 
Thus, the lowest observed H$\alpha$ flux indicates the
detection limit. In the work by \cite{skillman2003a}
this detection limit 
is about $4\cdot 10^{-16}$~erg~cm$^{-2}$~s$^{-1}$ 
(H {\sc ii} region ESO 348-G9~No.2). In the work
by \citet{miller1994a} the lowest measured 
H$\alpha$ flux in M81 group dwarf galaxies is about 
$6.5 \cdot 10^{-16}$~erg~cm$^{-2}$~s$^{-1}$ (IC 2574~MH~135). 
Assuming that each star cluster forms its own H {\sc ii} region
created by all ionising stars formed in this star cluster, a mass-luminosity
relation (Fig. \ref{fig_mass_luminosity}) between the embedded star cluster mass and the averaged produced
H$\alpha$ luminosity within the star formation epoch $\delta t$ 
can be constructed as described in Sec. \ref{ionising_photons}.
This can be well approximated by
\begin{equation}
  \label{equ_mass_luminosity}
  y = (x + 34.55)\;(1-1/\exp(0.27 x^2+0.46 x+0.32))\;\;\;,
\end{equation}
with $y = \log_{10}(L_{\mathrm{H}\alpha}/\mathrm{erg\;s}^{-1})$
and $x= \log_{10}(M_\mathrm{ecl}/\mathrm{M}_\odot)$. For comparison
the Orion Nebula cluster with a total mass of about 1800~M$_\odot$
\citep{hillenbrand1998a} and an observed  H$\alpha$ luminosity of 
10$^{37}$~erg~s$^{-1}$ \citep{kennicutt1984a} and 30~Doradus with a
total mass of about  272000~M$_\odot$  \citep{selman1999a}      
and an observed H$\alpha$ luminosity of 
$1.5\cdot 10^{40}$~erg~s$^{-1}$ \citep{kennicutt1984a} are marked, lying
well on the mass-H$\alpha$ luminosity relation. 

The Sculptor dwarf irregular galaxy SC~24, for example, 
has  only one observable H {\sc ii}-region. 
Its flux is $8\cdot 10^{-15}$~erg~cm$^{-2}$~s$^{-1}$ \citep{skillman2003a} 
being just  above the detection limit. 
Given the distance of  SC~24 of about 2.14~Mpc  the H$\alpha$ luminosity of
the most luminous H {\sc  ii}-region in SC~24 is $2.5\cdot 10^{35}$~erg~s$^{-1}$.
Given our mass-luminosity relation (Fig.~\ref{fig_mass_luminosity}) 
the corresponding star cluster stellar mass is about 280~M$_\odot$. 
Eq. \ref{eqn_m_ecl_max_sfr} then determines the current SFR to be about
$5.0 \cdot 10^{-4}$~M$_\odot$~yr$^{-1}$ in good agreement with the
Standard-IGIMF Scenario rather than the classical one 
(Tab. \ref{tab_sculptor-dIrr}). The classical SFR of 
$2.1\cdot 10^{-6}$~M$_\odot$~yr$^{-1}$ implies that only 1~M$_\odot$
of stellar material will have formed within 1~Myr which is by two
orders of magnitude inconsistent with the expected cluster mass 
(280~M$_\odot$). However, our SFR of $5.8\cdot 10^{-4}$ means that 
580~M$_\odot$ must have formed within 1~Myr showing an internal 
consistency of our picture. 

\begin{figure}
  \plotone{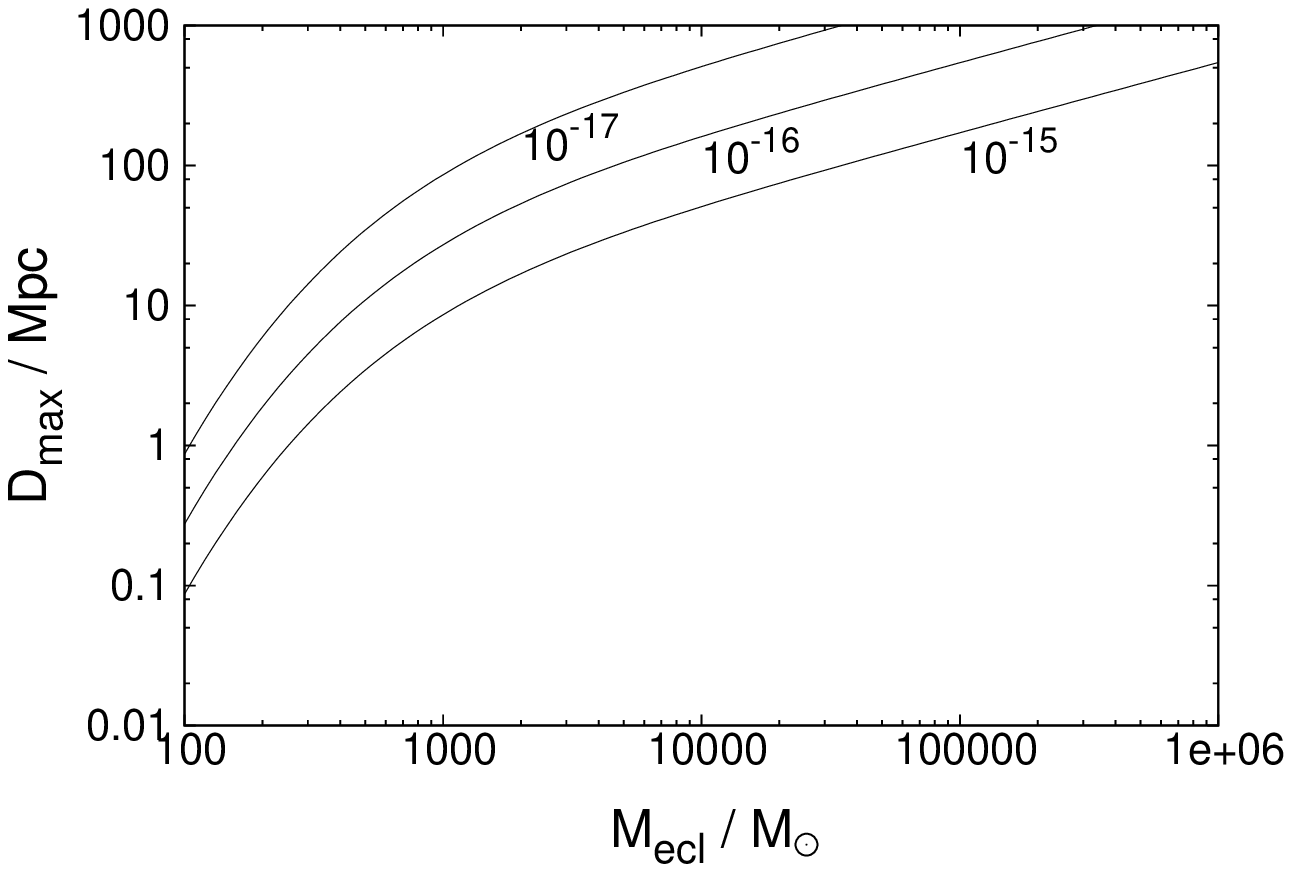}
  \caption{Relation between the stellar mass of embedded star clusters and
  the maximum distance of the star cluster within which 
  the star cluster can be detected with H$\alpha$ observations
  in the case of three different 
  detection limits: 10$^{-15}$, 10$^{-16}$ and 10$^{-17}$ erg~s$^{-1}$~cm$^{-2}$.}
  \label{fig_invisible_Halpha}
\end{figure}

An H {\sc ii}-region with the luminosity L$_{\mathrm{H}\alpha}$ in a
galaxy at a distance $D$ produces an H$\alpha$ flux of
\begin{equation}
j = \frac{L_{\mathrm{H}\alpha}}{4 \pi D^2}\;\;\;.
\end{equation}
Using the mass-H$\alpha$ luminosity relation
(Eq.~\ref{equ_mass_luminosity})
and a detection limit, $j_\mathrm{limit}$, a relation between
the star cluster mass and the maximum distance, $D_\mathrm{max}$,
within which the star cluster can be observed, can be 
constructed (Fig. \ref{fig_invisible_Halpha}).  

If the Sculptor dwarf galaxy SC~24 were located more distant than
2.3~Mpc then SF in SC~24 would not have been detected. 
Therefore very moderate SF can be overseen.
The effect of H$\alpha$-dark  star formation on the cosmological SFR 
due to a possible large
number of dwarf galaxies with low SFRs
needs further study.

\section{Conclusion}

Using the integrated galaxial initial mass function (IGIMF) instead of 
an invariant IMF
for calculating the produced H$\alpha$-luminosity for a given 
galaxy-wide SFR
we revise the linear $L_{\mathrm{H}\alpha}$-SFR relation by 
\citet{kennicutt1994a}. The order of magnitude of the deviation
of the IGIMF-$L_{\mathrm{H}\alpha}$-SFR relation from the linear 
$L_{\mathrm{H}\alpha}$-SFR relation is independent 
of the explicit choice of the IGIMF model.

Because the fraction of massive stars determined by the IGIMF
is always less than given by the underlying invariant IMF of individual star
clusters, SFRs are always underestimated when using linear
$L_\mathrm{H\alpha}$-SFR relations. The linearity 
between the H$\alpha$ luminosity and the SFR is broken, too. Therefore,
the SFRs of normal disk galaxies with an H$\alpha$ luminosity
of about 10$^{40}$--10$^{42}$~erg~s$^{-1}$ may be underestimated
by a factor of 4 up to 50. 
For dwarf irregular galaxies having H$\alpha$ luminosities between
10$^{-36}$ and 10$^{-39}$~erg~s$^{-1}$ the underestimation of the SFRs
vary by a factor of at least 20 to 160 for the Minimal Scenario.
In the  Maximal Scenario the underestimation of SFRs of dwarf galaxies
vary between a factor of 800  and 5000. 
Evolutionary  parameters of dwarf galaxies such as the gas depletion time scale
are effected by the same order of magnitude. 
In fact, we find the IGIMF
formulation to imply a linear dependence of the total SFR on the total
galaxy gas mass and therefore the gas depletion time scale to be constant, 
a few Gyr, and independent of the total galaxy gas mass. 

If this finding is correct it imposes a problem for the theoretical 
  understanding of galactic star formation 
  explaining the apparent low 
  star formation efficiency of small galaxies
  \citep[e.g.][]{kaufmann2007a}.

Observational evidence confirming the IGIMF scenario is becoming
available (Hoversten \& Glazebrook, 2007) but 
  further observational work is needed. Additionally, more
  consistency checks with different methods of deriving SFRs should be
  done in future.

Further  work is required to construct reliable
conversions of IGIMF-fluxes into other pass bands. 
\citet{kennicutt1998a} mentioned that H$\alpha$ measurements become 
unreliable for determining SFRs in regions with gas densities higher
than 100~M$_\odot$~pc$^{-2}$ as present in star burst galaxies. 
An extension of the IGIMF-model to construct a relation between SFRs
and FIR luminosities is therefore clearly necessary.
\acknowledgments
We thank Ivo Saviane for stimulating discussions which encouraged
us to write this paper.  
\bibliographystyle{mn2e}
\bibliography{imf,OB-star,star-formation,ONC,cmf,stellar-evolution,stellar-spectra,submitted,sfh,astro-ph}

\begin{thebibliography}{}

\bibitem[\protect\citeauthoryear{{Bonnell}, {Vine} \& {Bate}}{{Bonnell}
  et~al.}{2004}]{bonnell2004a}
{Bonnell} I.~A.,  {Vine} S.~G.,    {Bate} M.~R.,  2004, \mnras, 349, 735

\bibitem[\protect\citeauthoryear{{Bressan}, {Fagotto}, {Bertelli} \&
  {Chiosi}}{{Bressan} et~al.}{1993}]{bressan1993a}
{Bressan} A.,  {Fagotto} F.,  {Bertelli} G.,    {Chiosi} C.,  1993, \aaps, 100,
  647

\bibitem[\protect\citeauthoryear{{Bruzual} \& {Charlot}}{{Bruzual} \&
  {Charlot}}{2003}]{bruzual2003a}
{Bruzual} G.,  {Charlot} S.,  2003, \mnras, 344, 1000

\bibitem[\protect\citeauthoryear{{Diehl}, {Halloin}, {Kretschmer}, {Lichti},
  {Sch{\"o}nfelder}, {Strong}, {von Kienlin}, {Wang}, {Jean}, {Kn{\"o}dlseder},
  {Roques}, {Weidenspointner}, {Schanne}, {Hartmann}, {Winkler} \&
  {Wunderer}}{{Diehl} et~al.}{2006}]{diehl2006a}
{Diehl} R.,  {Halloin} H.,  {Kretschmer} K.,  {Lichti} G.~G.,
  {Sch{\"o}nfelder} V.,  {Strong} A.~W.,  {von Kienlin} A.,  {Wang} W.,  {Jean}
  P.,  {Kn{\"o}dlseder} J.,  {Roques} J.-P.,  {Weidenspointner} G.,  {Schanne}
  S.,  {Hartmann} D.~H.,  {Winkler} C.,    {Wunderer} C.,  2006, \nat, 439, 45

\bibitem[\protect\citeauthoryear{{Figer}}{{Figer}}{2005}]{figer2005a}
{Figer} D.~F.,  2005, \nat, 434, 192

\bibitem[\protect\citeauthoryear{{Hillenbrand} \& {Hartmann}}{{Hillenbrand} \&
  {Hartmann}}{1998}]{hillenbrand1998a}
{Hillenbrand} L.~A.,  {Hartmann} L.~W.,  1998, \apj, 492, 540

\bibitem[\protect\citeauthoryear{{Hoversten} \& {Glazebrook}}{{Hoversten} \&
  {Glazebrook}}{2007}]{hoversten2007a}
{Hoversten} E.,  {Glazebrook} K.,  2007, \apj, submitted (preprint)

\bibitem[\protect\citeauthoryear{{Hurley}, {Pols} \& {Tout}}{{Hurley}
  et~al.}{2000}]{hurley2000a}
{Hurley} J.~R.,  {Pols} O.~R.,    {Tout} C.~A.,  2000, \mnras, 315, 543

\bibitem[\protect\citeauthoryear{{Kaufmann}, {Wheeler} \& {Bullock}}{{Kaufmann}
  et~al.}{2007}]{kaufmann2007a}
{Kaufmann} T.,  {Wheeler} C.,    {Bullock} J.~S.,  2007, ArXiv e-prints, 706

\bibitem[\protect\citeauthoryear{{Kennicutt}
  Jr.}{{Kennicutt}}{1983}]{kennicutt1983a}
{Kennicutt} Jr. R.~C.,  1983, \apj, 272, 54

\bibitem[\protect\citeauthoryear{{Kennicutt}
  Jr.}{{Kennicutt}}{1984}]{kennicutt1984a}
{Kennicutt} Jr. R.~C.,  1984, \apj, 287, 116

\bibitem[\protect\citeauthoryear{{Kennicutt}
  Jr.}{{Kennicutt}}{1998}]{kennicutt1998a}
{Kennicutt} Jr. R.~C.,  1998, \apj, 498, 541

\bibitem[\protect\citeauthoryear{{Kennicutt} Jr., {Tamblyn} \&
  {Congdon}}{{Kennicutt} et~al.}{1994}]{kennicutt1994a}
{Kennicutt} Jr. R.~C.,  {Tamblyn} P.,    {Congdon} C.~E.,  1994, \apj, 435, 22

\bibitem[\protect\citeauthoryear{{Koen}}{{Koen}}{2006}]{koen2006a}
{Koen} C.,  2006, \mnras, 365, 590

\bibitem[\protect\citeauthoryear{{K{\"o}ppen}, {Weidner} \&
  {Kroupa}}{{K{\"o}ppen} et~al.}{2007}]{koeppen2007a}
{K{\"o}ppen} J.,  {Weidner} C.,    {Kroupa} P.,  2007, \mnras, 375, 673

\bibitem[\protect\citeauthoryear{{Kroupa}}{{Kroupa}}{2001}]{kroupa2001a}
{Kroupa} P.,  2001, \mnras, 322, 231

\bibitem[\protect\citeauthoryear{{Kroupa}}{{Kroupa}}{2002}]{kroupa2002a}
{Kroupa} P.,  2002, Science, 295, 82

\bibitem[\protect\citeauthoryear{{Kroupa}, {Tout} \& {Gilmore}}{{Kroupa}
  et~al.}{1993}]{kroupa1993a}
{Kroupa} P.,  {Tout} C.~A.,    {Gilmore} G.,  1993, \mnras, 262, 545

\bibitem[\protect\citeauthoryear{{Kroupa} \& {Weidner}}{{Kroupa} \&
  {Weidner}}{2003}]{weidner2003a}
{Kroupa} P.,  {Weidner} C.,  2003, \apj, 598, 1076

\bibitem[\protect\citeauthoryear{{Larsen}}{{Larsen}}{2000}]{larsen2000a}
{Larsen} S.~S.,  2000, \mnras, 319, 893

\bibitem[\protect\citeauthoryear{{Larsen}}{{Larsen}}{2002a}]{larsen2002b}
{Larsen} S.~S.,  2002a, in {Geisler} D.,  {Grebel} E.~K.,   {Minniti} D.,  eds,
  IAU Symposium {Open, Massive and Globular Clusters -- Part of the Same
  Family?}.
pp 421--+

\bibitem[\protect\citeauthoryear{{Larsen}}{{Larsen}}{2002b}]{larsen2002a}
{Larsen} S.~S.,  2002b, \aj, 124, 1393

\bibitem[\protect\citeauthoryear{{Larsen} \& {Richtler}}{{Larsen} \&
  {Richtler}}{2000}]{larsen2000b}
{Larsen} S.~S.,  {Richtler} T.,  2000, \aap, 354, 836

\bibitem[\protect\citeauthoryear{{Lee}, {Gibson}, {Flynn}, {Kawata} \&
  {Beasley}}{{Lee} et~al.}{2004}]{lee2004a}
{Lee} H.-c.,  {Gibson} B.~K.,  {Flynn} C.,  {Kawata} D.,    {Beasley} M.~A.,
  2004, \mnras, 353, 113

\bibitem[\protect\citeauthoryear{{Lejeune}, {Cuisinier} \& {Buser}}{{Lejeune}
  et~al.}{1997}]{lejeune1997a}
{Lejeune} T.,  {Cuisinier} F.,    {Buser} R.,  1997, \aaps, 125, 229

\bibitem[\protect\citeauthoryear{{Lejeune}, {Cuisinier} \& {Buser}}{{Lejeune}
  et~al.}{1998}]{lejeune1998a}
{Lejeune} T.,  {Cuisinier} F.,    {Buser} R.,  1998, \aaps, 130, 65

\bibitem[\protect\citeauthoryear{{Ma{\'{\i}}z Apell{\'a}niz}, {Walborn},
  {Morrell}, {Niemela} \& {Nelan}}{{Ma{\'{\i}}z Apell{\'a}niz}
  et~al.}{2006}]{maiz_apellaniz2006a}
{Ma{\'{\i}}z Apell{\'a}niz} J.,  {Walborn} N.~R.,  {Morrell} N.~I.,  {Niemela}
  V.~S.,    {Nelan} E.~P.,  2006, ArXiv Astrophysics e-prints

\bibitem[\protect\citeauthoryear{{Maschberger} \& {Kroupa}}{{Maschberger} \&
  {Kroupa}}{2007}]{maschberger2007a}
{Maschberger} T.,  {Kroupa} P.,  2007, \mnras, 379, 34

\bibitem[\protect\citeauthoryear{{McKee} \& {Williams}}{{McKee} \&
  {Williams}}{1997}]{mckee1997a}
{McKee} C.~F.,  {Williams} J.~P.,  1997, \apj, 476, 144

\bibitem[\protect\citeauthoryear{{Meynet} \& {Maeder}}{{Meynet} \&
  {Maeder}}{2003}]{meynet2003a}
{Meynet} G.,  {Maeder} A.,  2003, \aap, 404, 975

\bibitem[\protect\citeauthoryear{{Miller} \& {Hodge}}{{Miller} \&
  {Hodge}}{1994}]{miller1994a}
{Miller} B.~W.,  {Hodge} P.,  1994, \apj, 427, 656

\bibitem[\protect\citeauthoryear{{Oey} \& {Clarke}}{{Oey} \&
  {Clarke}}{2005}]{oey2005a}
{Oey} M.~S.,  {Clarke} C.~J.,  2005, \apjl, 620, L43

\bibitem[\protect\citeauthoryear{{Pflamm-Altenburg} \&
  {Kroupa}}{{Pflamm-Altenburg} \& {Kroupa}}{2006}]{pflamm_altenburg2006a}
{Pflamm-Altenburg} J.,  {Kroupa} P.,  2006, \mnras, 373, 295

\bibitem[\protect\citeauthoryear{{Scalo}}{{Scalo}}{1986}]{scalo1986a}
{Scalo} J.~M.,  1986, Fundamentals of Cosmic Physics, 11, 1

\bibitem[\protect\citeauthoryear{{Schaller}, {Schaerer}, {Meynet} \&
  {Maeder}}{{Schaller} et~al.}{1992}]{schaller1992a}
{Schaller} G.,  {Schaerer} D.,  {Meynet} G.,    {Maeder} A.,  1992, \aaps, 96,
  269

\bibitem[\protect\citeauthoryear{{Selman}, {Melnick}, {Bosch} \&
  {Terlevich}}{{Selman} et~al.}{1999}]{selman1999a}
{Selman} F.,  {Melnick} J.,  {Bosch} G.,    {Terlevich} R.,  1999, \aap, 347,
  532

\bibitem[\protect\citeauthoryear{{Selman} \& {Melnick}}{{Selman} \&
  {Melnick}}{2005}]{selman2005a}
{Selman} F.~J.,  {Melnick} J.,  2005, \aap, 443, 851

\bibitem[\protect\citeauthoryear{{Skillman}, {C{\^o}t{\'e}} \&
  {Miller}}{{Skillman} et~al.}{2003}]{skillman2003a}
{Skillman} E.~D.,  {C{\^o}t{\'e}} S.,    {Miller} B.~W.,  2003, \aj, 125, 593

\bibitem[\protect\citeauthoryear{{Stahler} \& {Palla}}{{Stahler} \&
  {Palla}}{2005}]{stahler2005a}
{Stahler} S.~W.,  {Palla} F.,  2005, {The Formation of Stars}.
The Formation of Stars, by Steven W.~Stahler, Francesco Palla, pp.~865.~ISBN
  3-527-40559-3.~Wiley-VCH , January 2005.

\bibitem[\protect\citeauthoryear{{Vanbeveren}}{{Vanbeveren}}{1982}]{vanbeveren%
1982a}
{Vanbeveren} D.,  1982, \aap, 115, 65

\bibitem[\protect\citeauthoryear{{Vanbeveren}}{{Vanbeveren}}{1983}]{vanbeveren%
1983a}
{Vanbeveren} D.,  1983, \aap, 124, 71

\bibitem[\protect\citeauthoryear{{Weidner} \& {Kroupa}}{{Weidner} \&
  {Kroupa}}{2004}]{weidner2004a}
{Weidner} C.,  {Kroupa} P.,  2004, \mnras, 348, 187

\bibitem[\protect\citeauthoryear{{Weidner} \& {Kroupa}}{{Weidner} \&
  {Kroupa}}{2005}]{weidner2005a}
{Weidner} C.,  {Kroupa} P.,  2005, \apj, 625, 754

\bibitem[\protect\citeauthoryear{{Weidner} \& {Kroupa}}{{Weidner} \&
  {Kroupa}}{2006}]{weidner2006a}
{Weidner} C.,  {Kroupa} P.,  2006, \mnras, 365, 1333

\bibitem[\protect\citeauthoryear{{Weidner}, {Kroupa} \& {Larsen}}{{Weidner}
  et~al.}{2004}]{weidner2004b}
{Weidner} C.,  {Kroupa} P.,    {Larsen} S.~S.,  2004, \mnras, 350, 1503

\bibitem[\protect\citeauthoryear{{Westera}, {Lejeune}, {Buser}, {Cuisinier} \&
  {Bruzual}}{{Westera} et~al.}{2002}]{westera2002a}
{Westera} P.,  {Lejeune} T.,  {Buser} R.,  {Cuisinier} F.,    {Bruzual} G.,
  2002, \aap, 381, 524

\end{thebibliography}

\end{document}